\documentclass[twocolumn,aps,superscriptaddress,prb,longbibliography]{revtex4-1}
\usepackage{graphicx}
\usepackage{amssymb}
\usepackage{amsmath}
\usepackage{bm}
\usepackage{color}

\def\Journal #1,#2,#3,#4#5#6#7{#1 {\bf #2}, #3 (#4#5#6#7)}

\def\Vec{\mathbf}
\def\GVec#1{\mbox{\boldmath $#1$}}

\begin{document}

\title{Moir\'{e} phonons in the twisted bilayer graphene\\
}
\author{Mikito Koshino}
\thanks{koshino@phys.sci.osaka-u.ac.jp}
\affiliation{Department of Physics, Osaka University,  Toyonaka 560-0043, Japan}

\author{Young-Woo Son}
\affiliation{Korea Institute for Advanced Study, Seoul 02455, Korea}

\begin{abstract}
We study the in-plane acoustic phonons in twisted bilayer graphenes
using the effective continuum approach.
We calculate the phonon modes by solving the continuum equation of motion
for infinitesimal vibration around the static relaxed state with triangular domain structure.
We find that the moir\'{e} interlayer potential only affects the in-plane asymmetric modes,
where the original linear dispersion is broken down into miniphonon bands separated by gaps,
while the in-plane symmetric modes with their linear dispersion are hardly affected.
The phonon wave functions of asymmetric modes are regarded as collective vibrations of 
the domain-wall network, and the low-energy phonon band structure can be qualitatively described 
by an effective moir\'{e}-scale lattice model.
\end{abstract}

\maketitle

\section{Introduction}
\label{sec_intro}

Twisted bilayer graphene (TBG), or a pair of graphene layers rotationally stacked on top of each other\cite{berger2006electronic,hass2007structural,hass2008multilayer,li2009observation,miller2010structural,luican2011single}, exhibits a variety of physical properties depending on the twist angle $\theta$.
In a small $\theta$, in particular, a long-period moir\'{e} pattern due to a slight lattice mismatch
folds the Dirac cone of graphene into a superlattice Brillouin zone
\cite{lopes2007graphene,mele2010commensuration,trambly2010localization,shallcross2010electronic,morell2010flat,bistritzer2011moirepnas,kindermann2011local,xian2011effects,PhysRevB.86.155449,moon2012energy,de2012numerical,moon2013opticalabsorption,weckbecker2016lowenergy},
and nearly-flat bands with extremely narrow band width 
emerge at some particular $\theta$'s called the magic angles.
The discovery of the superconductivity and strongly correlated insulating state 
in the magic-angle TBGs \cite{cao2018unconventional,cao2018mott,yankowitz2019tuning}  
has attracted enormous attention to this system.

While the early theoretical studies on TBG simply assumed a stack of rigid graphene layers without deformation,
the actual TBG spontaneously relaxes to the energetically favorable lattice structure.
There the in-plane distortion maximizes the area of AB stacking (graphite's Bernal stacking) 
to form a triangular domain pattern \cite{popov2011commensurate, brown2012twinning,lin2013ac,alden2013strain,
van2015relaxation,dai2016twisted, jain2016structure,nam2017lattice,carr2018relaxation, lin2018shear,yoo2019atomic,walet2019lattice},
and at the same time the out-of-plane distortion leads to a corrugation of graphene layers \cite{uchida2014atomic,van2015relaxation,lin2018shear}.
The electronic band structure is also significantly affected by the lattice relaxation
\cite{nam2017lattice, lin2018shear, koshino2018maximally, yoo2019atomic,walet2019lattice,crucial2019lucignano}.
A similar structural relaxation was also observed in moir\'{e} superlattice of graphene and hexagonal boron-nitride 
(h-BN)\cite{Woods2014GhBN},
and the effects on the band structure were investigated
\cite{san2014spontaneous,San-Jose2014GhBN,jung2015origin}.

The lattice deformation of the moir\'{e} superlattice is expected to significantly influence the phonon properties.
In the intrinsic graphene, the low-energy phonon spectrum is composed of transverse acoustic (TA) 
modes and longitudinal acoustic (LA) modes which have linear dispersions,
and the out-of-plane flexural phonons (ZA) with a quadratic dispersion 
\cite{gruneis2002determination,suzuura2002phonons,yan2008phonon}.
In the AB-stacked bilayer graphene, the phonon modes of the top layer and the bottom layer are 
coupled to form layer-symmetric and asymmetric modes \cite{yan2008phonon,nika2017phonons}.
The phonon spectrum of TBG was calculated by fully taking account of all the atoms
in the moir\'{e} unit cells \cite{jiang2012acoustic,cocemasov2013phonons,choi2018strong,angeli2019valley}, 
where the phonon density of states is found to be close to that of the regular AB-stacked bilayer graphene
insensitively to the twist angle.
On the other hand, it is naturally expected that the moir\'{e} interlayer potential 
would cause superlattice zone-folding and miniband generation in the phonon spectrum.
%Some more detailed analysis of phonon modes is required to understand those effects.
So far such zone-folding effect was argued for graphene h-BN superlattice \cite{felix2018thermal}
and very recently for the optical branch  of TBG \cite{angeli2019valley}.
The low frequency phonon minibands in the TBGs 
are not thoroughly understood although they may have important roles 
for low-energy transport and collective phenomena \cite{choi2018strong,wu2018theory, wu2019phononinduced}.

In this paper, we study the in-plane acoustic phonons in TBG 
and investigate the effects of the moir\'{e} superlattice structure on the phonon spectrum.
The calculation is based on the continuum approach
using the elastic theory and the registry-dependent interlayer potential,
which was used to obtain the relaxed lattice structure with a triangular domain pattern \cite{nam2017lattice}.
We derive the phonon modes by solving the continuum equation of motion 
for infinitesimal vibration around the static relaxed state.
We find that the moir\'{e} interlayer potential couples only to the in-plane asymmetric modes
(i.e., the top and bottom layers slide in the opposite directions parallel to the graphene layers),
while the in-plane symmetric modes (which slide in the same direction) are hardly affected.
In the in-plane asymmetric modes, the original linear dispersion 
is broken into miniphonon bands separated by gaps,
where the phonon wave functions are regarded as collective vibrations of 
the nano-scale triangular lattice of AB- and BA-stacking domains.
The formation of miniphonon bands and gaps is more pronounced in lower twist angles.
We find that the phonon band structure in the low twist angles becomes nearly invariant 
when renormalized by the energy scale inversely proportional to the moir\'{e} superlattice period.
The universal behavior of the  moir\'{e}  phonon bands 
can be understood by an effective moir\'{e}-scale lattice model.

%%%%%

\section{Methods}
\label{sec_methods}

 \begin{figure}
\begin{center}
%\leavevmode\includegraphics[width=0.9\hsize]{fig_lattice.eps}
\leavevmode\includegraphics[width=0.9\hsize]{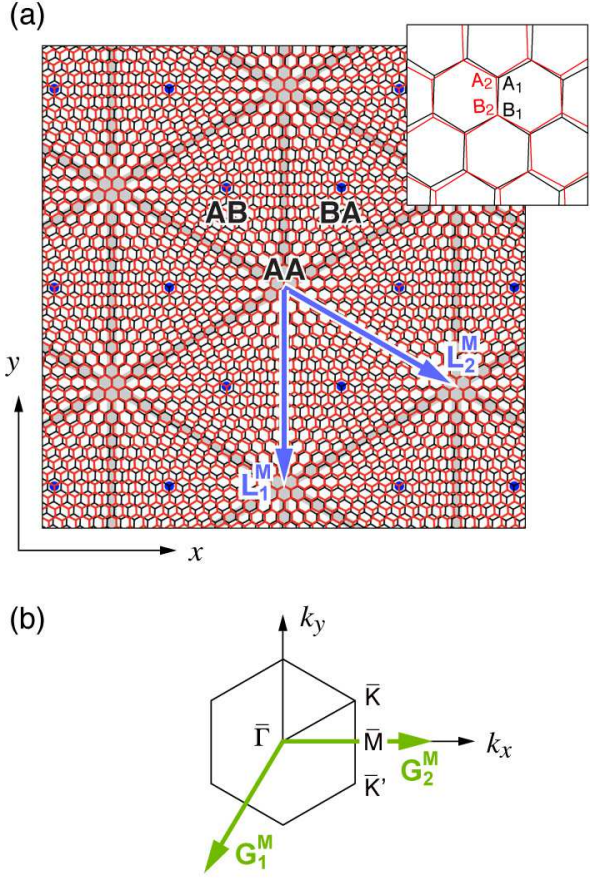}
\end{center}
\caption{
(a) Non-relaxed atomic structure of TBG with $\theta = 3.89^\circ$.
AA spots are located at the crossing points of the grid lines.
(b) First Brillouin zone of TBG.
}
\label{fig_lattice}
\end{figure}

We consider a TBG lattice with a relatively small twist angle $\theta$ lower than a few degree,
as shown in Fig.\ \ref{fig_lattice}(a).
Here we define the geometry of TBG 
by starting from  perfectly overlapping honeycomb lattices
and rotating the layer 1 and 2 by $-\theta/2$ and $+\theta/2$, respectively.
We define $\Vec{a}_1^{(0)} = a(1,0)$ and $\Vec{a}_2^{(0)} = a(1/2,\sqrt{3}/2)$ 
as the primitive lattice vectors of the initial bilayer graphene before the rotation, 
where $a \approx 0.246\,\mathrm{nm}$ is the graphene's lattice constant. 
Then the  lattice vectors of layer 1 are given by $\Vec{a}_i =R(- \theta/2)\Vec{a}_i^{(0)}$
and those of layer 2 are given by  $\tilde{\Vec{a}}_i =R(\theta/2) \Vec{a}_i^{(0)}$ ($i=1,2$),
 where $R(\phi)$ represents the rotation matrix by angle $\phi$.
The corresponding reciprocal lattice vectors, $\Vec{a}^*_i$ and $\tilde{\Vec{a}}^*_i$ 
for layer 1 and 2, respectively,
are defined by $\Vec{a}_i \cdot \Vec{a}^*_j = 2\pi\delta_{ij}$ and 
$\tilde{\Vec{a}}_i \cdot \tilde{\Vec{a}}^*_j = 2\pi\delta_{ij}$.

%We define the primitive lattice vectors of layer 1 as $\textbf{a}_1 = a(1,0)$,  
%$\textbf{a}_2 = a(1/2,\sqrt{3}/2)$, and those of layer 2 as $\tilde{\textbf{a}}_i =R(\theta) \textbf{a}_i\, (i=1,2)$,
%where $a = 0.246$ nm is graphene's lattice constant and $R(\theta)$ is the two-dimensional rotation matrix. 
%The reciprocal lattice vectors of layer 1 are given by $\Vec{a}^*_1 = 2\pi/a (1,-1/\sqrt{3})$ 
%and $\Vec{a}^*_2 = 2\pi/a(0,2/\sqrt{3})$,
%and those of layer 2 by  $\tilde{\Vec{a}}^*_i =R(\theta) \Vec{a}^*_i$ ($i=1,2$).    

When the twist angle is small, the slight mismatch of the lattice periods of two layers 
gives rise to a long-period moir\'{e} beating pattern.
The primitive lattice vector of the moir\'{e} superlattice $\Vec{L}_i^{\rm M}$ is 
given by \cite{nam2017lattice}
\begin{equation}
%  \Vec{L}_i^{\rm M} = (1-R^{-1})^{-1} \Vec{a}_i\quad (i=1,2).
 \Vec{L}_i^{\rm M} =[1-R(-\theta)]^{-1} \Vec{a}_i\quad (i=1,2).
\end{equation}
The lattice constant $L_{\rm M} = | \Vec{L}_1^{\rm M}|=| \Vec{L}_2^{\rm M}|$ becomes
 \begin{equation}
       L_{\rm M} =  \frac{a}{2\sin (\theta/2)}.
       \end{equation}
The corresponding moir\'{e} reciprocal lattice vectors 
satisfying $\Vec{G}^{\rm M}_i\cdot\Vec{L}_j^{\rm M} = 2\pi\delta_{ij}$
are written as
\begin{equation}
%  \Vec{G}_i^{\rm M} = (1-R)\, \textbf{a}^*_i
  \Vec{G}_i^{\rm M} = [1-R(\theta)]\, \textbf{a}^*_i
  =  \textbf{a}^*_i - \tilde{\textbf{a}}^*_i. \quad (i=1,2).
\end{equation}
The first Brillouin zone defined by $\Vec{G}^{\rm M}_i$ is shown in Fig.\ \ref{fig_lattice}(b).
The local structure of TBG approximates the nonrotated bilayer graphene with the in-plane
translation. It is characterized by the interlayer sliding vector, or the shift of layer 2's atom at $\Vec{r}$, 
measured from its counterpart on layer 1.
Without the lattice relaxation, the interlayer sliding vector is given by 
\begin{equation}
%\bm{\delta}_0 (\textbf{r}) = \textbf{r} - \textbf{r}_0 = (1 - {R}^{-1}) \textbf{r}.
\bm{\delta}_0 (\textbf{r}) = \textbf{r} - \textbf{r}_0 =[1-R(-\theta)] \textbf{r}.
\label{eq_delta_0}
\end{equation}

Now we introduce the in-plane lattice vibration specified by the time-dependent displacement vector,
$\textbf{u}^{(l)}(\textbf{r},t)$ for layer $l =1,2$.
The interlayer sliding vector under the deformation is 
 \begin{equation}
\bm{\delta}(\textbf{r},t) = \bm{\delta}_0 (\textbf{r}) + \textbf{u}^{(2)}(\textbf{r},t) -  \textbf{u}^{(1)}(\textbf{r},t).
\label{eq_delta_TBG}
 \end{equation}
Here we concentrate on the in-plane displacement,
while the qualitative argument of out-of-plane phonon modes will be presented in Sec.\ \ref{sec_disc}.
The inter-layer binding energy of TBG is written as
\begin{eqnarray}\label{Eq:U_B_TBG}
U_B  &=& \int V[\bm{\delta} (\textbf{r},t)]   \text{d}^2 \textbf{r},
\end{eqnarray}
where $V[\bm{\delta}]$ is the interlayer binding energy per area of nonrotated bilayer graphene
with the sliding vector $\bm{\delta}$ \cite{nam2017lattice}.
It can be approximately written as
\begin{equation}
V[\bm{\delta}]  = \sum^3_{j= 1} 2 V_0 \cos [\textbf{a}^*_j \cdot \bm{\delta}],
\label{eq_v_TBG}
\end{equation}
where $\textbf{a}^*_3 = -\textbf{a}^*_1 - \textbf{a}^*_2$.
The function takes the maximum value $6V_0$ at AA stacking ($\bm{\delta}=0$)
and the minimum value $-3V_0$ at AB and BA stacking.
The difference between the binding energies of AA and AB/BA structure 
is $9V_0$ per area, and this amounts to $\Delta  \epsilon = 9V_0 S_G/ 4$ per atom
where $S_G$ is the area of graphene's unit cell.
In the following calculation, we use $\Delta \epsilon = 0.0189$ (eV/atom) as a typical value \cite{Lebedeva2011,popov2011commensurate}.
By using Eqs.\ (\ref{eq_delta_TBG})	 and (\ref{eq_v_TBG}), we have
\begin{equation}
V[\bm{\delta} (\textbf{r},t)]   = \sum^3_{j= 1}  
2V_0 \cos [ \textbf{G}_j^{\rm M} \cdot \textbf{r} + \textbf{a}^*_j\cdot(\textbf{u}^{(2)} -\textbf{u}^{(1)}) ],
\end{equation}
where $ \textbf{G}^{\rm M}_3 = - \textbf{G}^{\rm M}_1 - \textbf{G}^{\rm M}_2$ and
 we used the relation $\textbf{a}^*_j \cdot \bm{\delta}_0(\textbf{r})=\textbf{G}_j^{\rm M} \cdot \textbf{r}$.

The elastic energy of strained TBG is expressed by \cite{suzuura2002phonons,San-Jose2014GhBN}
\begin{align}
 U_E = \sum_{l=1}^2 \int \frac{1}{2} \left\{ (\lambda +\mu ) (u_{xx}^{(l)} + u_{yy}^{(l)})^2 \right. \nonumber\\
 \left.   +\mu \left[ ( u_{xx}^{(l)} - u_{yy}^{(l)})^2 + 4(u_{xy}^{(l)})^2 \right] \right\} d^2 \textbf{r},
 \label{Eq:U_E_TBG}
\end{align}
 %\begin{eqnarray}\label{Eq:U_E_TBG}
% U_E =\sum_{l=1}^2 \int \limits_{S_M} \frac{1}{2} 
 %\left[ 2\mu \text{Tr} (\textbf{u}_l^2) +\lambda (\text{Tr} (\textbf{u}_l))^2 \right] \text{d}^2 \textbf{r},		
 %\end{eqnarray}
 where % $S_{\rm M}= (\sqrt{3}/2)L_{\rm M}^2$ is the area of moir\'e unit cell, 
 $\lambda \approx 3.25$ eV/$\text{\AA}^2$ and
 $\mu \approx 9.57$ eV/$\text{\AA}^2$ are graphene's 
 $\text{Lam}\acute{\text{e}}$  factors \cite{zakharchenko2009finite,jung2015origin}, 
and $u^{(l)}_{ij} = (\partial_i u_j^{(l)} + \partial_j u_i^{(l)})/2$ is  the strain tensor.
Lastly, the kinetic energy due to the motion of the lattice is expressed as
\begin{align}
 T= \sum_{l=1}^2 \int 
 \frac{\rho}{2} \left[\dot{u}_{x}^{(l)2} + \dot{u}_{y}^{(l)2}\right] d^2 \textbf{r},
 \label{Eq:T_TBG}
\end{align}
where $\rho= 7.61\times 10^{-7}$ kg/m$^2$ is the area density of single-layer graphene,
and $\dot{u_i}$ represents the time derivative of $u_i\,(i=x,y)$.

The Lagrangian of the system is given by $L=T-(U_E+U_B)$
as a functional of $\textbf{u}^{(l)}(\textbf{r})$.
We define $\textbf{u}^{\pm} = \textbf{u}^{(2)} \pm \textbf{u}^{(1)}$ and rewrite 
$L$ as a functional of $\textbf{u}^{\pm}$.
The Euler-Lagrange equations for  $\textbf{u}^{-}$ read
\begin{multline}\label{Eq:Euler1}
\frac{1}{2} (\lambda +\mu) \left( \frac{\partial^2u^{-}_x}{\partial x^2} 
+\frac{\partial^2 u^{-}_y}{\partial x \partial y} \right) 
+ \frac{\mu}{2} \left( \frac{\partial^2u^{-}_x}{\partial x^2} 
+\frac{\partial^2 u^{-}_x}{\partial y^2} \right)\\
+ \sum_{j=1}^3 2 V_0 \sin [\textbf{G}_j^{\rm M} \cdot \textbf{r} + \textbf{a}^*_j \cdot \textbf{u}^{-}]a^*_{jx} = 
\frac{1}{2} \rho \ddot{u}^{-}_x,
\end{multline}
\begin{multline}\label{Eq:Euler2}
\frac{1}{2} (\lambda +\mu) \left( \frac{\partial^2u^{-}_y}{\partial y^2} +\frac{\partial^2 u^{-}_x}{\partial x \partial y} \right) 
+ \frac{\mu}{2} \left( \frac{\partial^2u^{-}_y}{\partial x^2} 
+\frac{\partial^2 u^{-}_y}{\partial y^2} \right)\\
+ \sum_{j=1}^3 2 V_0 \sin [\textbf{G}_j^{\rm M} \cdot \textbf{r} + \textbf{a}^*_j \cdot \textbf{u}^{-}] a^*_{jy} = 
\frac{1}{2} \rho \ddot{u}^{-}_y,
\end{multline}
where $a^*_{jl}\, (l=x,y)$ is the $l$-component of $\Vec{a}^*_j$.
The equation of motion for  $\textbf{u}^{+}$ is given by replacing $\textbf{u}^{-}$ with $\textbf{u}^{+}$ and 
removing all the terms including $V_0$  in Eqs.\ (\ref{Eq:Euler1}) and (\ref{Eq:Euler2}).
Here the potential terms with $V_0$ vanish because $U_B$ is not dependent on $\textbf{u}^{+}$
and then the force term $\partial U_B/ \partial \textbf{u}^{+}$ is zero.
Therefore, the mode $\textbf{u}^{+}$ is just equivalent to the original acoustic phonon of  graphene.

For $\textbf{u}^{-}$, we consider a small vibration around the static equilibrium state, or
\begin{align}
\textbf{u}^{-} (\Vec{r},t) =  \textbf{u}^{-}_0 (\Vec{r}) + \delta\textbf{u}^{-} (\Vec{r},t).
\end{align}
Here $\textbf{u}^{-}_0 (\Vec{r})$ is the static solution of Eqs.\ (\ref{Eq:Euler1}) and (\ref{Eq:Euler2}),
i.e., the optimized relaxed state to minimize $U_B+U_E$,
and $\delta\textbf{u}^{-} (\Vec{r},t)$ is the perturbational excitation around $\textbf{u}^{-}_0$.
%which is assumed to much smaller than the graphene's atomic constant $a$.
We define the Fourier components as
 \begin{align}
  & \textbf{u}_0^{-} (\textbf{r})  = 
\sum_{\textbf{G}} \textbf{u}^{-}_{0,\textbf{G}} e^{i\textbf{G}\cdot \textbf{r}},
\label{eq_u0_fourier}
 \\
& \delta\textbf{u}^{-} (\textbf{r},t)  = 
e^{-i \omega t}
\sum_{\textbf{q}}\delta\textbf{u}^{-}_{\textbf{q}} e^{i\textbf{q}\cdot \textbf{r}},
\label{eq_delta_u_fourier}
 \\
& \sin \left[\textbf{G}_j^{\rm M} \cdot \textbf{r} + \textbf{a}^*_j \cdot \textbf{u}^{-}_0(\textbf{r})\right] 
= \sum_{\textbf{G}} f^j_{\textbf{G}} e^{i \textbf{G} \cdot \textbf{r}}, 
\label{eq_cos_fourier}
\\
& \cos \left[\textbf{G}_j^{\rm M} \cdot \textbf{r} + \textbf{a}^*_j \cdot \textbf{u}^{-}_0(\textbf{r})\right] 
= \sum_{\textbf{G}} h^j_{\textbf{G}} e^{i \textbf{G} \cdot \textbf{r}},
\label{eq_sin_fourier}
\end{align}
where $\textbf{G} = m \textbf{G}^{\rm M}_1 + n \textbf{G}^{\rm M}_2$ are moir\'{e} reciprocal vectors
and $\omega$ is the phonon frequency.

From Eqs.\ (\ref{Eq:Euler1}) and (\ref{Eq:Euler2}),
the equation for the static solution $\textbf{u}_0^{-}$ is given by \cite{nam2017lattice}
 \begin{align} 
&        \textbf{u}^{-}_{0, \textbf{G}} =  
\sum_{j =1}^3  4V_0 f^j_{\textbf{G}} \hat{K}_{\textbf{G}}^{-1} \textbf{a}^*_j,
\notag\\
&        \hat{K}_{\textbf{G}}=\left( {\begin{array}{cc}
        (\lambda+2\mu) G_x^2 + \mu G_y^2 &  (\lambda + \mu)G_x G_y \\
        (\lambda +\mu) G_x G_y & (\lambda + 2\mu)G_y^2 + \mu G_x^2
        \end{array}} \right) .
        \label{Eq:E-LMatrix}
 \end{align}
The equation of motion for the dynamical perturbation part is written as
  \begin{align} 
&       \rho\, \omega^2  \delta \textbf{u}^{-}_{\textbf{G}+\textbf{q}} =  
 \hat{K}_{\textbf{G}+\textbf{q}}  \delta \textbf{u}^{-}_{\textbf{G}+\textbf{q}}
 - 4 V_0 \sum_{\textbf{G}'} \hat{V}_{\textbf{G}-\textbf{G}'} \delta \textbf{u}^{-}_{\textbf{G}'+\textbf{q}},
\notag\\
&        \hat{V}_{\textbf{G}}=
\sum_{j =1}^3  h^j_{\textbf{G}} 
\begin{pmatrix}
a^*_{jx}a^*_{jx} & a^*_{jx}a^*_{jy}\\
a^*_{jx}a^*_{jy} & a^*_{jy}a^*_{jy}
\end{pmatrix}.
\label{eq_delta_u}
\end{align}
 
We first derive the static solution $\textbf{u}_0^{-}$
by solving a set of the self-consistent equations,
Eqs. (\ref{eq_u0_fourier}), (\ref{eq_cos_fourier}) and (\ref{Eq:E-LMatrix}),
by numerical iterations \cite{nam2017lattice}.
Using the obtained $\textbf{u}_0^{-}$,
we solve the eigenfunction Eq.\ (\ref{eq_delta_u}),
to obtain eigenvalues $\omega^2$ and
the phonon modes $\delta \textbf{u}^{-}$ as a function of $\Vec{q}$ in the 
moir\'{e} Brillouin zone.
Throughout the calculation, we set the $k$-space cut-off 
for the Fourier components of $\textbf{u}_0^{-}$ and $\delta \textbf{u}^{-}$, 
which is sufficiently large for convergence.

%The number of the relavant harmonics in $\textbf{u}^{-}_{\textbf{q}}$ 
%is roughly characterized by a dimensionless parameter
%\begin{eqnarray}
%\eta = \sqrt{\frac{V_0}{\lambda + \mu}} \,\frac{L_{\rm M}}{a}.
%%\eta = \frac{V_0 L_{\rm M}^2}{(\lambda + \mu) a^2}.
%\label{Eq:Eta_2D}
%\end{eqnarray}

%%%%%%%

\begin{figure*}
\begin{center}
\leavevmode\includegraphics[width=0.9\hsize]{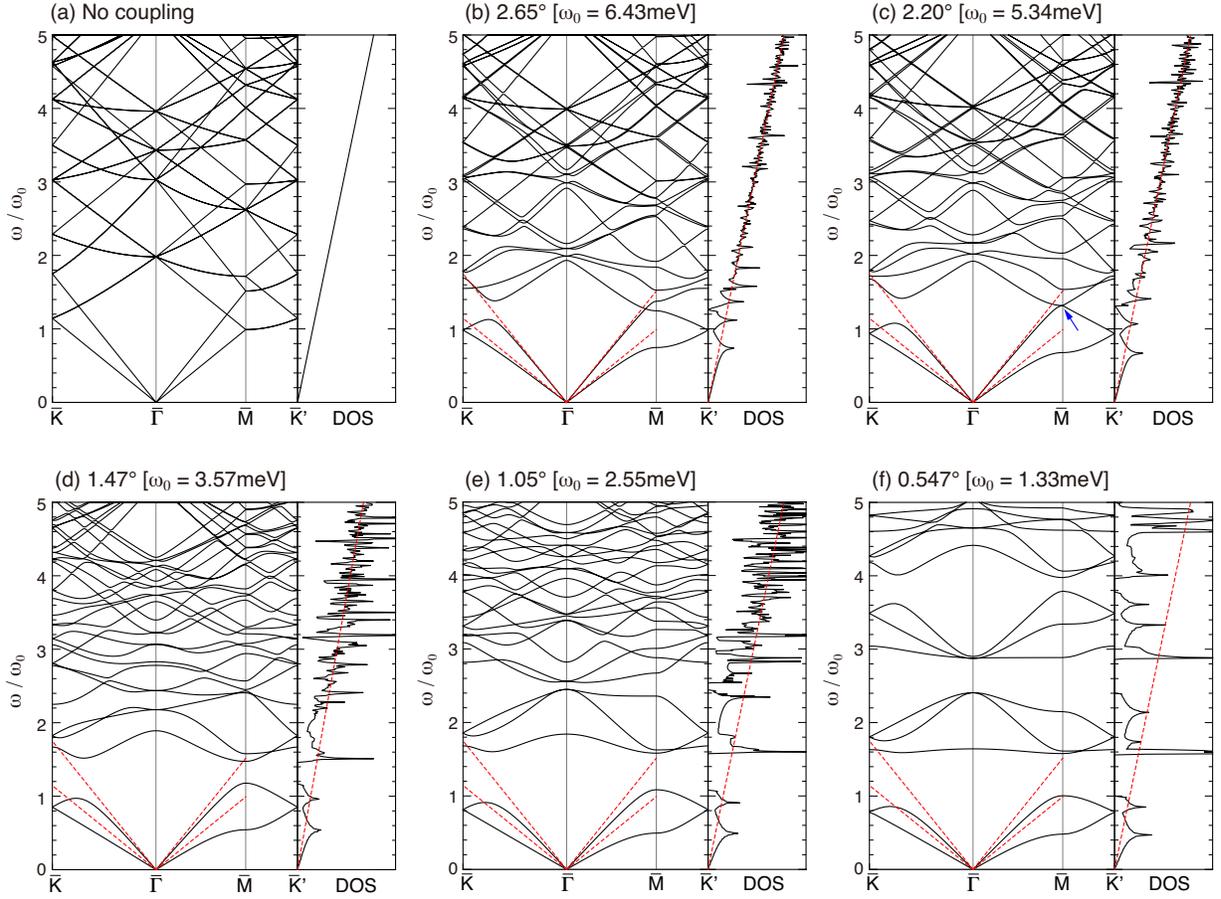}
\end{center}
\caption{
Phonon dispersions of in-plane asymmetric modes ($\Vec{u}^-$) in various twist angles $\theta$.
The horizontal axis is scaled by the superlattice Brillouin zone size $\propto 1/L_M$, and the vertical axis is scaled by 
$\omega_0 \propto 1/L_M$ (see the text).
(a) Phonon dispersion when the moir\'{e} interlayer coupling is absent.
}
\label{fig_phonon_band}
\end{figure*}

\begin{figure*}
\begin{center}
%\leavevmode\includegraphics[width=0.8\hsize]{fig_phonon_band3D.eps}
\leavevmode\includegraphics[width=0.8\hsize]{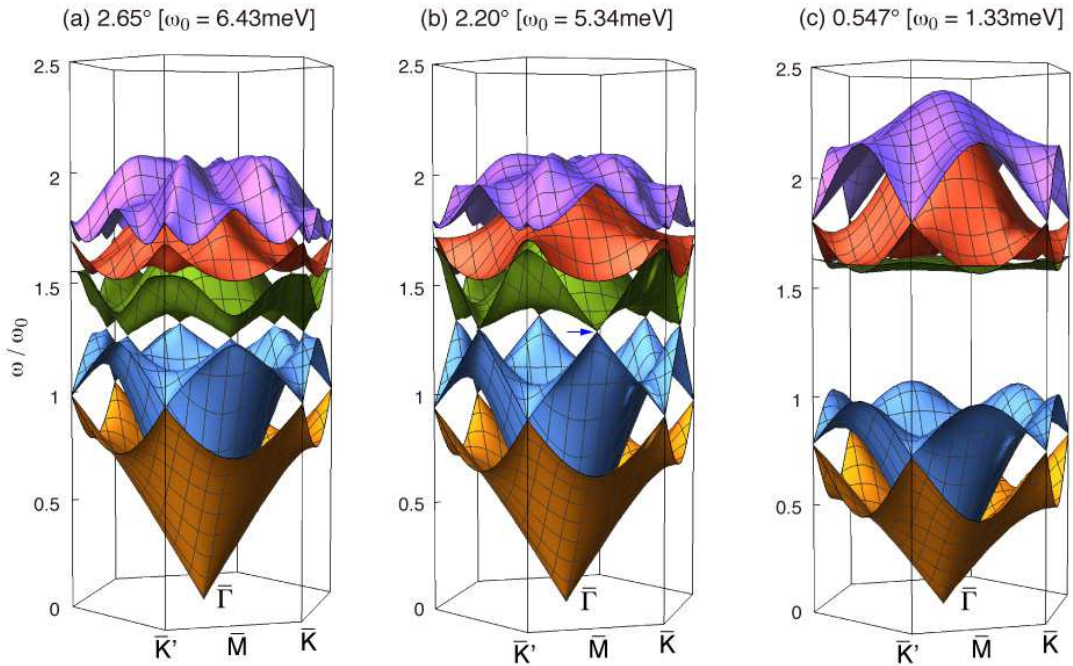}
\end{center}
\caption{
Two-dimensional dispersions of the lowest five modes
of  in-plane asymmetric vibration ($\Vec{u}^-$)  in (a) $\theta = 2.65^\circ$,
(b) 2.20$^\circ$ and (c) 0.547$^\circ$.
}
\label{fig_phonon_band3D}
\end{figure*}

\begin{figure}
\begin{center}
%\leavevmode\includegraphics[width=1.\hsize]{fig_phonon_wave.eps}
\leavevmode\includegraphics[width=1.\hsize]{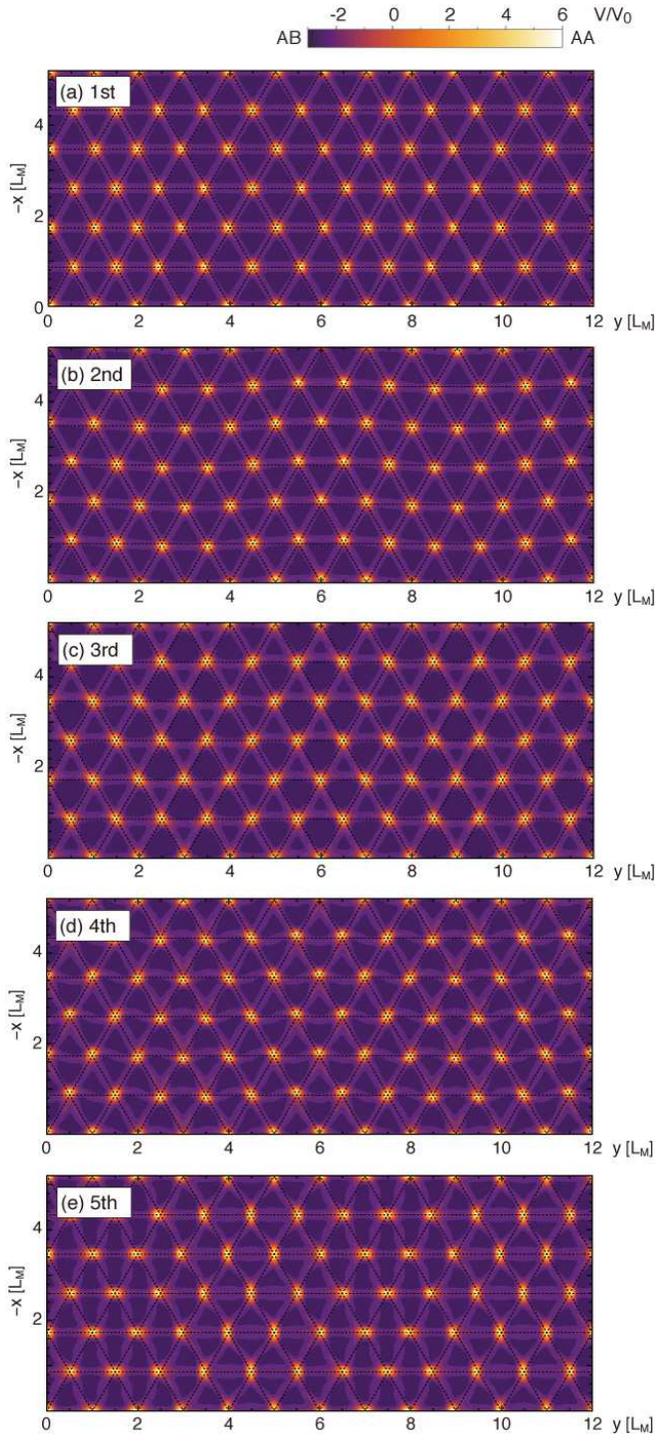}
\end{center}
\caption{
Phonon wave functions of the lowest five modes
at $\Vec{q} = (0,2\pi/(6L_M))$ in TBG of $\theta=1.05^\circ$.
The $y$ axis is taken as the horizontal axis,
and the color code represents the local binding energy $V[\GVec{\delta}(\Vec{r},t)]$ at a certain $t$.
Dashed lines indicate the superlattice without vibrations.
The corresponding band structure is shown in Fig.\ \ref{fig_phonon_band}(d).
}
\label{fig_phonon_wave}
\end{figure}

\begin{figure}
\begin{center}
\leavevmode\includegraphics[width=0.9\hsize]{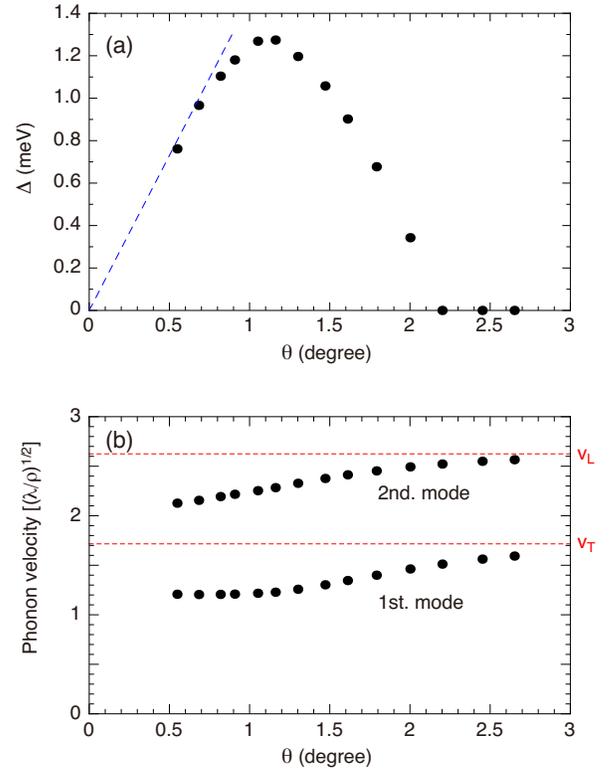}
\end{center}
\caption{
(a) Width of the gap (in meV) between the second and the third branches
of $\Vec{u}^-$  phonon modes, plotted against the twist angle. The blue dashed line is a guide to the eye
indicating the linear dependence on $\theta$.
(b) Group velocities (in units of $\sqrt{\lambda/\rho}$, long wavelength limit) of the first and the second 
$\Vec{u}^-$  phonon modes, plotted against the twist angle. 
Horizontal dashed lines indicate the velocities of the transverse ($v_T$) and longitudinal ($v_L$) acoustic phonons 
in single layer graphene.
}
\label{fig_gap_vs_theta}
\end{figure}

\begin{figure}
\begin{center}
%\leavevmode\includegraphics[width=0.6\hsize]{fig_schem_domain.eps}
\leavevmode\includegraphics[width=0.6\hsize]{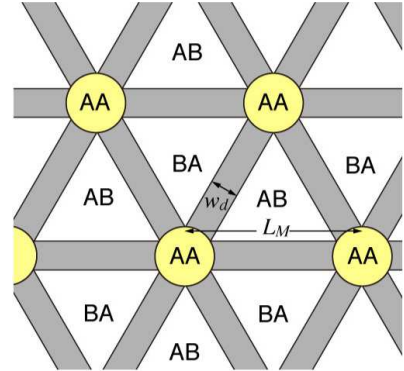}
\end{center}
\caption{
Schematics of the domain structure in TBG.
The shaded region indicates the domain walls separating AB and BA regions.
}
\label{fig_schem_domain}
\end{figure}

\begin{figure}
\begin{center}
\leavevmode\includegraphics[width=1.\hsize]{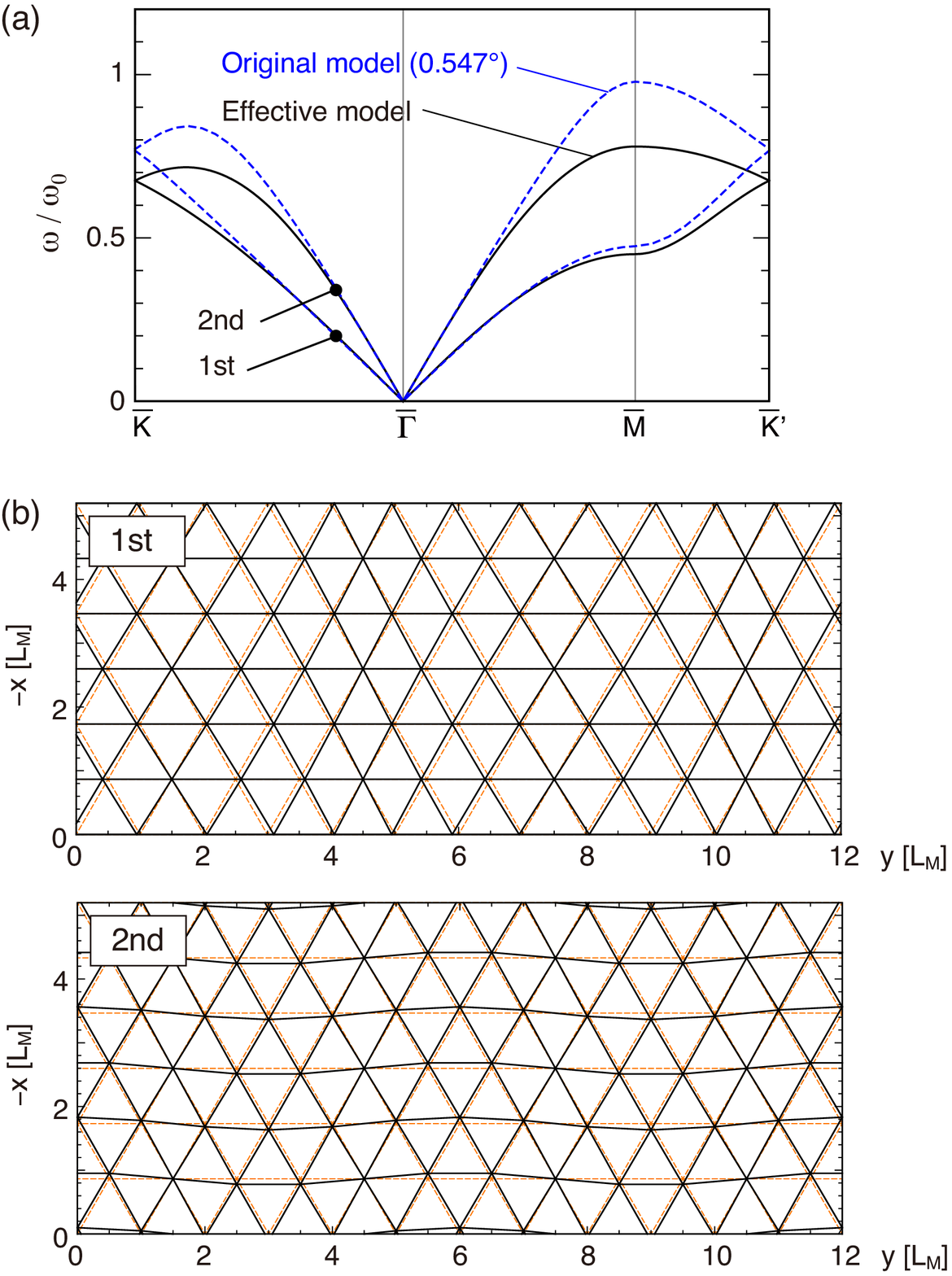}
\end{center}
\caption{
(a) Phonon dispersions of the effective lattice model (black solid)
and the original model for $\theta = 0.547^\circ$ from Fig.\ \ref{fig_phonon_band}(f) (blue dashed).
(b) Phonon vibration of the effective lattice
for the first and the second modes at $\Vec{q} = (0,2\pi/(6L_M))$, which are indicated by dots in panel (a). 
Orange dashed lines indicate the lattice without vibrations.
}
\label{fig_effective_model}
\end{figure}

%%%%%

\section{Phonon modes}
\label{sec_phonon}

Figure \ref{fig_phonon_band} shows the phonon dispersions 
of in-plane asymmetric modes ($\Vec{u}^-$) calculated for various twist angle $\theta$'s.
Here the horizontal axis is the scaled by the superlattice Brillouin zone size
($\propto 2\pi /L_M$) and the vertical axis is scaled by 
\begin{align}
\omega_0 = \sqrt{\frac{\lambda}{\rho}} \frac{2\pi}{L_M},
\end{align}
where $\sqrt{\lambda/\rho}$ is the characteristic velocity scale for the acoustic phonons in graphene.
Figure \ref{fig_phonon_band}(a) is the phonon dispersion when the moir\'{e} interlayer coupling is absent.
This is equivalent to the folded dispersion of intrinsic TA and LA phonons of graphene,
which are $\omega(\Vec{q}) = v_T q$ and $v_L q$, respectively,
where $v_T = \sqrt{\mu/\rho}$ and $v_L = \sqrt{(\lambda + 2\mu)/\rho}$.\cite{suzuura2002phonons}
Figure \ref{fig_phonon_band}(a) is independent of the twist angle,
since both the horizontal and vertical axes are normalized by units proportional to $1/L_M$.
The same dispersions are also indicated by red dashed lines in Figs.\ \ref{fig_phonon_band}(b)-(f).
Note that the phonon band structures of in-plane symmetric modes ($\Vec{u}^+$) remain intact 
and ungapped. So, they are not drawn in this figure or hereafter.

For $\Vec{u}^-$, we clearly see that the original linear dispersions of graphene's acoustic phonons are
reconstructed into superlattice minibands.
In the lower twist angles, in particular, 
we observe that the lowest two branches are separated by a  gap from the rest of the spectrum.
Figure \ref{fig_phonon_band3D} shows the two-dimensional dispersion of the lowest five $\Vec{u}^-$ modes
in $\theta = 2.65^\circ$, 2.20$^\circ$ and 0.547$^\circ$,
the dispersions of which along the high-symmetric lines are shown in Fig.\ \ref{fig_phonon_band}(b), (c) and (f),
respectively.
In $2.65^\circ$, the second and the third phonon bands stick at the six wave points in the Brillouin zone. 
At the critical angle $\theta_c \approx 2.20^\circ$,
these touching points annihilate in pairs at three $\bar{M}$ points
[indicated by a blue arrow in Fig.\ \ref{fig_phonon_band}(c) and Fig.\ \ref{fig_phonon_band3D}(b)],
and the full gap opens in $\theta < \theta_c$.
In the lowest twist angle 0.547$^\circ$, 
we also see that the spectral gaps exist between the fifth and sixth modes and  the eighth and ninth modes
[Fig. \ref{fig_phonon_band}(f)].

In Fig.\ \ref{fig_phonon_wave},  we plot the phonon wave functions of the lowest five $\Vec{u}^-$ modes
at $\Vec{q} = (0,2\pi/(6L_M))$ ($=(1/4)\bar{\Gamma} \bar{\textrm{K}}$)  in $\theta=1.05^\circ$.
Here the $y$ axis is taken as the horizontal axis,
and the color code represents the local binding energy $V[\GVec{\delta}(\Vec{r},t)]$ at a certain $t$.
Note that, in making these plots, we assume finite vibrational amplitudes for the purpose of illustration,
although the phonon modes are actually obtained for infinitesimal displacements.
The lowest mode (a) and the second lowest mode (b) are regarded as
the longitudinal and transverse acoustic modes of an effective lattice
with “atoms” at the AA sites and “bonds” between them defined by the AB-BA domain walls.
In the limit of zero interlayer coupling [Fig. \ref{fig_phonon_band}(a)],
the longitudinal (transverse) mode of the effective lattice becomes the transverse 
(longitudinal) mode of intrinsic graphene.
Here the transverse and longitudinal directions become opposite
in graphene lattice and the effective lattice, because the movement of graphene atoms
and the corresponding movement of the AA spots are at 90 degree to each other;
e.g., if the graphene layers are shifted in the $x$ direction, the AA spots move along the $y$ direction.

Importantly, we notice that the low-energy part of the phonon dispersion (scaled by $\omega_0$)
is almost identical at $\theta = 1.05^\circ$ and $0.547^\circ$ [Fig.\ \ref{fig_phonon_band}(e) and (f)],
indicating that it is converging to the universal structure in the low twist angle limit.
This is somewhat surprising because the energy scale of the 
interlayer interacting potential $V_0$ (a constant) is increasing relative to the reference 
scale $\omega_0 (\propto \theta)$ as $\theta$ decreases, 
so that one may naively expect the phonon band gap  should increase relative to $\omega_0$ as well.
In Fig.\ \ref{fig_gap_vs_theta}(a), we plot the width of the gap (in meV) between the second and the third branches
as a function of the twist angle.
In decreasing $\theta$, it takes the maximum around $\theta \sim 1.1^\circ$, and then
it turns to decrease and approach the linear behavior in $\theta$,
indicating the gap size in units of $\omega_0$ is converging to a constant value.
Figure \ref{fig_gap_vs_theta}(b) shows the twist angle dependence of
the phonon group velocities $\partial \omega/\partial q |_{q\to 0}$ 
for the first and the second $\Vec{u}^-$  modes. 
In large twist angles, they become the velocities of the transverse ($v_T$) and longitudinal ($v_L$) acoustic phonon modes of single layer graphene (indicated by the horizontal dashed lines), as argued above.
In decreasing the angle, the phonon velocities decrease as a result of the miniband formation,
and they eventually approach constant values in the low angle limit.

This universal feature of the scaled phonon dispersion can be understood by considering the domain-wall structure
in the relaxed TBG, which is schematically illustrated as Fig.\ \ref{fig_schem_domain}.
Here triangular AB and BA regions are separated by the domain walls (gray regions)
with AA stacked spots at vertices.
It was shown that the characteristic width of the domain walls is almost independent of the twist angle \cite{alden2013strain}, and it is given by \cite{nam2017lattice}
\begin{align}
w_d \approx \frac{a}{4}\sqrt{\frac{\lambda+\mu}{V_0}},
\label{eq_w_d}
\end{align}
which is about 5 nm.
The domain pattern becomes clear in low twist angles such that $L_M \gg w_d$, or $\theta$ lower than 
about 2 degrees.
We first qualitatively understand the scale of $w_d$ by the following order estimation.
The area of the domain wall per moir\'{e} unit cell is given by $\sim w_d L_M $ with the numerical factor neglected.
The total interlayer binding energy (relative to the AB / BA stacking) per moir\'{e} unit cell 
is then $U_B \sim V_0 (w_d L_M)$.
The elastic energy is also concentrated to the domain-wall regions 
and it is given by $U_E \sim \lambda (\partial u_i/\partial x_j)^2 (w_d L_M)$.
Here we took $\lambda$ as a representative scale for the elastic constant,
since $\lambda$ and $\mu$ are in the same order of magnitude.
In the domain wall, the strain tensor $\partial u_i/\partial x_j$ is of the order 
of $a/w_d$ since the atomic shift $u_i$ changes by about $a$ inside the domain wall.
The relaxed state is given by the condition $U_E \sim U_B$,
and this gives $w_d \sim a \sqrt{\lambda/V_0}$, which has the same order of magnitude as Eq.\ (\ref{eq_w_d}).

Now let us consider an oscillation of the moir\'{e} lattice around the relaxed state.
We consider a simple excitation such as the Fig. \ref{fig_phonon_wave}(a) and (b),
and assume the in-plane displacement of the AA positions (`lattice points') at $\Vec{R}$ 
is $\tilde{\Vec{u}}(\Vec{R})$.
During the oscillation, the binding energy and elastic energy are shown to be
approximately proportional to the total length of the domain walls.
In Figs.\ \ref{fig_phonon_wave}(a) and (b), we actually see that the domain walls (`bonds')
are elongated (or shortened), and here the graphene's atomic lattice is not stretched in the same direction,
but the area of the same local atomic structure is just increased,
so that the total energy change is proportional to the domain-wall length.
The change of the total domain-wall length from the relaxed state is 
of the order of $\tilde{u}^2/L_M$ per a moir\'{e} unit cell,
noting that the linear term in $\tilde{u}$ can be negative and positive place by place and vanishes in total.
Therefore, the changes in the binding energy and the elastic energy are given by
$\delta U_B \sim V_0 w_d \, \tilde{u}^2/L_M$ and $\delta U_E \sim \lambda (a/w_d)^2 w_d \, \tilde{u}^2/L_M$,
which are of the same order because $V_0 \sim \lambda (a/w_d)^2$ as argued above.
The kinetic energy $T$ is estimated by considering the atomic motion.
Here the change of the moir\'{e} lattice $\tilde{u}$ is actually caused by a change of the interlayer atomic shift $\Vec{u}^-$  around the domain wall. Here we have a relation $\tilde{u} \sim (w_d / a)  \delta u^-$,
because a change of the atomic lattice of the order of $a$ is magnified to
a change of moir\'{e} lattice by $w_d$. 
Considering that the atoms are oscillating only around the domain-wall regions,
the kinetic energy per moir\'{e} unit cell is given by
$T \sim \rho (\delta u^-)^2 \omega^2 (w_dL_M)  = \rho  (a / w_d)^2 \, \tilde{u}^2 \omega^2 (w_dL_M)$ 
where $\omega$ is the oscillation frequency.
By assuming the kinetic energy $T$ and the potential energy $U_E$ and $U_B$ are of
the same order of magnitude, we end up with
\begin{align}
\omega \sim \sqrt{\frac{\lambda}{\rho}} \frac{1}{L_M},
\end{align}
which correctly gives the order of $\omega_0$.

Based on the above argument, we can construct a simple effective model
to describe the lowest two modes.
We consider a triangular lattice composed of masses (corresponding to the AA spots) 
and bonds (domain walls) connecting the neighboring masses.
We assume that the bonds are always straight 
and the energy of each bond is proportional to its length.
We consider a vibration around the perfect triangular lattice with the lattice constant $L_M$,
assuming that the total area of the system is fixed.
If the in-plane displacement of the mass is given by $\tilde{\Vec{u}}(\Vec{R})$,
the length change of the bonds summed over the whole system is given by
\begin{align}
\Delta L_{\rm bond} = \sum_{\Vec{R}} \sum_{i=1}^3
\frac{1}{2L_M}
\left[
|\Delta \tilde{\Vec{u}}^{(i)}(\Vec{R})|^2 
-\Bigl(\Delta \tilde{\Vec{u}}^{(i)}(\Vec{R}) \cdot \frac{\Vec{L}^M_i}{L_M} \Bigr)^2
\right]
\end{align}
where $\Delta \tilde{\Vec{u}}^{(i)}(\Vec{R}) = \tilde{\Vec{u}}(\Vec{R} + \Vec{L}^M_i) -\tilde{\Vec{u}}(\Vec{R})$,
the vector $\Vec{R} = m \Vec{L}^M_1 + n \Vec{L}^M_2$ runs over the lattice points,
and we define $\Vec{L}^M_3 =\Vec{L}^M_2- \Vec{L}^M_1$.
The total energy of the bonds is given by $\tilde{U} = \alpha V_0 w_d \Delta L_{\rm bond}$,
where $\alpha$ is a numerical factor of the order of 1 to match the energy scale with the original model.
By the Fourier transformation 
$\tilde{\Vec{u}}(\Vec{R}) = \sum_\Vec{q} \tilde{\Vec{u}}_{\Vec{q}} e^{i \Vec{q}\cdot\Vec{R}}$,
it is written as
\begin{equation}
\tilde{U} =  \frac{1}{2} \sum_\Vec{q} 
\tilde{\Vec{u}}_{-\Vec{q}}^T 
\hat{D}(\Vec{q}) \tilde{\Vec{u}}_{\Vec{q}},
\end{equation}
where $\hat{D}(\Vec{q}) $ is the dynamical matrix defined by
\begin{equation}
D_{\mu \nu}(\Vec{q}) = \sum_{i=1}^3 \frac{\alpha V_0 w_d}{L_M} 
\Biggl(
2 \sin \frac{\Vec{q}\cdot\Vec{L}^M_i}{2}
\Biggr)^2
\Biggl[
\delta_{\mu\nu} - \frac{(\Vec{L}^M_i)_\mu (\Vec{L}^M_i)_\nu}{L_M^2}
\Biggr],
\label{eq_D}
\end{equation}
where $\mu,\nu = x,y$. 
The kinetic energy is given by 
$\tilde{T} = \frac{1}{2}\sum_\Vec{q} 
M \dot{\tilde{\Vec{u}}}_{-\Vec{q}} \cdot \dot{\tilde{\Vec{u}}}_{\Vec{q}}$, where
 $M = \rho (a/w_d)^2 w_d L_M$ is the effective mass. 
Finally the equation of motion is given by 
$M  \ddot{\tilde{\Vec{u}}}_{\Vec{q}} = - \hat{D}(\Vec{q}) \tilde{\Vec{u}}_{\Vec{q}}$.
It is straight forward to see that the equation can be transformed to a dimensionless form
by scaling the frequency by $\omega_0$.
Figure \ref{fig_effective_model}(a) plots the phonon dispersion of this model,
where the blue dashed line represents the original phonon dispersion of $\theta = 0.547^\circ$.
Here we choose $\alpha = 2$ for the best fitting in the long wave-length region near $\bar{\Gamma}$.
We see that the effective model qualitatively reproduces the whole dispersion relation of the original model
for the first and second low-energy modes,
and we can also show that the phonon wave functions are also correctly reproduced in the whole 
Brillouin zone.
Figure \ref{fig_effective_model}(b) illustrates
the wave functions of the first and the second modes at $\Vec{q} =  (0,2\pi/(6L_M))$,
which agree with Figs.\ \ref{fig_phonon_wave}(a) and (b), respectively.

The vibration of the moir\'{e} lattice is different from that in an ordinary spring-mass model,
in that the potential energy of a bond (a domain wall) is proportional 
to its length, but not to the squared length.
This is directly related to the important property that the moir\'{e} longitudinal mode [Fig.\ \ref{fig_phonon_wave}(a)]
has a lower frequency than the moir\'{e} transverse mode [Fig. \ref{fig_phonon_wave}(b)] 
unlike in the usual elastic system.
In the ordinary spring-mass model, the strain energy of the moir\'{e} 
longitudinal mode as in Fig.\ \ref{fig_phonon_wave}(a)
is mostly contributed from the bonds parallel to the wave vector 
(horizontal bonds in this figure), where the length change gives the tensile energy cost.
In the present system, however, the horizontal bonds do not change the total energy
because elongated bonds and shortened bonds just cancel in the total length,
while the diagonal bonds contribute to a relatively small increase in the total length.
In the effective lattice model given by Eq.\ (\ref{eq_D}), it is straight forward to check that
the phonon frequency of the transverse mode is $\sqrt{3}$ times as large as that of the longitudinal mode
in the long wave-length limit. In the original model, the ratio of the second band frequency to the first 
is about 1.8  near the $\Gamma$ point, indicating that the effective model is valid.

The linear dependence of energy for elongation of bonds implies that there is a constant force for all stretching modes,
while such modes are prohibited in the present calculation because the moir\'{e} unit cell size $L_M$ is fixed.
In the original lattice model for TBG, the all stretching mode
corresponds to the overall interlayer rotation to reduce the twist angle $\theta$,
where the moir\'{e} unit cell monotonically expands.
The total energy (per area) then decreases
because the area of the AB regions increases in proportion to $L_M^2$,
and it eventually dominates over the area of the domain walls (with the width fixed to $w_d$),
which is just in proportion to $L_M$.
In experiments, a TBG with a finite twist angle can also exist as a meta-stable phase,
where certain external constraints caused by 
local pinning mechanisms should prohibit the global rotation.
The calculation fixing moir\'{e} unit cell size should be justified in such a situation.

As a final remark, we notice that the third lowest phonon band in Fig.\ \ref{fig_phonon_band} becomes flatter 
when the twist angle is reduced,  
resulting in a significant enhancement of phonon density of states as $\theta$ decreases.
The corresponding wave function in Fig.\ \ref{fig_phonon_wave}(c)
can be viewed as a `domain-breathing' mode
where triangular AB and BA domains expand and shrink in opposite phases.
One may also view this mode as an in-plane optical mode of the moir\'{e} lattice.
Here we see that the positions of the corner points of triangles are fixed during the oscillation.
This means that the oscillation of a triangle does not affect that of neighboring triangles,
so the system can be approximately viewed as independent oscillators.
We presume that it is the reason for the flat band nature, i.e. the constant frequency independent of the wave number.
The fourth and the fifth modes [Fig.\ \ref{fig_phonon_wave}(d) and (e)]
exhibit complicated wave patterns including the bending of the domain walls
and also the distortion of AA spots.

\section{Considerations on the flexural phonon}
\label{sec_disc}

While the current model only considered the in-plane phonons,
the actual TBG also has the flexural phonon modes vibrating in the out-of-plane direction.
Here we consider the effect of moir\'{e} interlayer coupling on the flexural phonons in TBG
by the following simple qualitative argument.
%We start from the non-distorted TBG in which two flat graphenes are stacked with interlayer spacing $d_{AB}$
%(the distance for AB-stacked bilayer),
%and consider three-dimenisional displacement $\Vec{u}^{(l)} = (u_x^{(l)},u_y^{(l)},u_z^{(l)})$
%from that state.
We consider three-dimenisional displacement $\Vec{u}^{(l)} = (u_x^{(l)},u_y^{(l)},u_z^{(l)})$
from the nondistorted configuration in which
two flat graphenes are stacked with interlayer spacing $d_{AB}$ (the distance for the AB-stacked bilayer).
It is known that the relaxed TBG is corrugated in such a way that
the spacing is minimum at AB/BA-stacking regions and maximum at AA-stacking regions
\cite{uchida2014atomic, van2015relaxation,lin2018shear}.
We can define the optimized interlayer spacing $d_{\rm opt}(\GVec{\delta})$ as a function
of the two-dimensional sliding vector $\GVec{\delta}(\Vec{r})$ [Eq.\ (\ref{eq_delta_TBG})].
The interlayer binding potential for the vertical displacement $(u^{(1)}_z, u^{(2)}_z)$
can then be written as a quadratic form
$V_\perp = \alpha [u^{-}_z + d_{AB} - d_{\rm opt}(\GVec{\delta})]^2$,
where $u^{-}_z = u^{(2)}_z - u^{(1)}_z$.
The total binding energy is a sum of $V_\perp$ and the in-plane part, Eq.\ (\ref{eq_v_TBG}).
The potential curvature $\alpha$ is shown to be nearly independent of the stacking structure, \cite{lee2008growth}
so we assume $\alpha$ is constant.
The effect of the moir\'{e} superlattice comes only from the potential term through 
$\GVec{\delta}=\GVec{\delta}(\Vec{r})$.
The kinetic part and elastic part are unchanged from the intrinsic graphene's and 
do not depend on  $\GVec{\delta}$.
In the equation of motion, we have the force term
$\partial V_\perp / \partial u_z = 2\alpha [u^{-}_z + d_{AB} - d_{\rm opt}(\GVec{\delta})]$,
and the optimized corrugated structure is given by $u^{-}_{0z} = d_{\rm opt}(\GVec{\delta}) - d_{AB}$.
When we consider an excitation from the optimized state $\Vec{u}^{-} =  \Vec{u}^{-}_{0}  +\delta \Vec{u}^{-}$,
the force term becomes $\partial V_\perp / \partial u_z = 2\alpha \delta u^{-}_z$,
and does not depend on $\GVec{\delta}$ anymore, i.e., the superlattice effect vanishes in the equation of motion 
for $\delta u^{-}_z$.

From these arguments, we presume that the flexural phonon does not exhibit well-pronounced miniband structure
unlike the in-plane asymmetric modes, 
and the corresponding phonon density of states remains mostly unchanged from the intrinsic graphene.
Actually the minigap formation was not observed in a phonon calculation 
fully including all the atomic sites in TBG with the registry-dependent interlayer interaction
\cite{choi2018strong,angeli2019valley}.
It implies that the flexural phonons remain almost intact,
considering that the density of states in the low-energy region of graphene is dominated by the flexural phonons
\cite{lindsay2010flexural}.
Rigorously speaking, the corrugated structure should cause some finite coupling between the in-plane modes and 
out-of-plane modes, and we leave the detailed study of the full three-dimensional phonon problem for future works.

\section{Conclusion}
\label{sec_conclusion}

We studied the in-plane acoustic phonons of low-angle TBGs
using the continuum approach.
We found that the moir\'{e} superlattice effect only affects the in-plane asymmetric modes
where the linear dispersion of the acoustic phonon is reconstructed into minibands separated by gaps,
while the in-plane symmetric modes remain intact.
The phonon wave functions for in-plane asymmetric modes can be regarded as vibrations of 
the triangular superlattice, and they are qualitatively understood by the effective lattice model of the moir\'{e} scale.

Our calculation shows that 50\% of the linear acoustic modes of the TBG are completely reconstructed
by the moir\'{e} superlattice coupling and their phonon velocities are significantly lowered.
Moreover, it is also shown that as the twist angle decreases, the characteristic flat phonon mode with diverging density of states is generated.
We expect that these effects should affect the thermal conductivity directly
and also phonon-related thermal phenomena.
The electron-phonon coupling between the flat-band electrons and the moir\'{e} phonons,
and its effect on the superconducting and correlating states, are also important problems.
We leave these issues for future study.

\section{Acknowledgments}

M. K. acknowledges fruitful discussions with Pablo Jarillo-Herrero, Philip Kim and  Debanjan Chowdhury. 
M. K. acknowledges the financial support of JSPS KAKENHI Grant Number JP17K05496. 
Y.-W.S. was supported by National Research Foundation of Korea (Grant No. 2017R1A5A1014862, SRC program: vdWMRC center).

%{\it Note added:}
%After completion of the present study, we have come to notice that
%a recent preprint which reports the full phonon calculations in TBG and 
%the coupling to the electronic states  \cite{angeli2019valley}

\bibliography{moire_phonon}

\end{document}